\documentstyle[twoside,fleqn,espcrc2,epsf]{article}

\newcommand\be{\begin{equation}}
\newcommand\ee{\end{equation}}

\newcommand\bea{\begin{eqnarray}}
\newcommand\eea{\end{eqnarray}}
\newcommand\half{{\textstyle{1\over2}}}
\newcommand\quarter{{\textstyle{1\over4}}}
\newcommand{\smfrac}[2]{{\textstyle{{#1}\over{#2}}}}

\newcommand{\E}[1]{e^{\textstyle {#1}}}
\def\da{^\dagger}
\def\d{\partial}
\def\n{{\mathbf n}}

\newcommand{\AmS}{{\protect\the\textfont2
  A\kern-.1667em\lower.5ex\hbox{M}\kern-.125emS}}

\title{Magnetic Monopole Content of Hot Instantons}

\author{R. C. Brower\address{Physics Department,
        Boston University, 590 Commonwealth Ave, Boston, MA 02215, USA}%
        \thanks{Talk presented by R.C.~Brower. Supported by the US DOE.
         Computations were done at the Theoretical
         Physics Computing Facility at Brown University.},
D. Chen$^{\rm b}$, J. Negele\address{Center for Theoretical Physics, 
Massachusetts Institute of Technology, Cambridge, MA 02139, USA},
K. Orginos\address{Physics Department, University of Arizona, Tucson, AZ 85721, USA} and  C-I Tan\address{Physics Department, Brown University, Providence, RI 02912, USA}}
       
\begin{document}

\begin{abstract}
We study the Abelian projection of an instanton in $R^3 \times S^1$ as
a function of temperature (T) and non-trivial holonomic twist
($\omega$) of the Polyakov loop at infinity. These parameters
interpolate between the circular monopole loop solution at $T=0$
and the static 't Hooft-Polyakov monopole/anti-monopole pair
at high temperature.
\end{abstract}

\maketitle

\section{INTRODUCTION}
Although many qualitative features of QCD are well described by a
vacuum state dominated by an instanton ``liquid', confinement appears
to be an exception \cite{lat98}. Instead, magnetic monopoles are
thought to be the crucial ingredient. This raises the question of how
magnetic degrees of freedom can be incorporated into (or reconciled
with) an instanton ``liquid''.  A recent step in this
direction was taken by Brower, Orginos and Tan
(BOT)\cite{BOT} who studied in detail the magnetic content of a single
isolated instanton, defining magnetic currents via the Maximally
Abelian (MA) projection. They found a marginally stable direction for
the formation of a monopole loop. Now with the
more general caloron solution of T. Kraan and P. van Baal
\cite{KvB}, and K. Lee and C. Lu \cite{LL}, this analysis can
be extended to an isolated SU(2) instanton at finite temperature ($T$)
with a non-trivial holonomy ($\omega$) for the Polyakov loop.
\begin{figure}[t]
\vspace{-0.1in}
\epsfxsize=\hsize
\epsffile{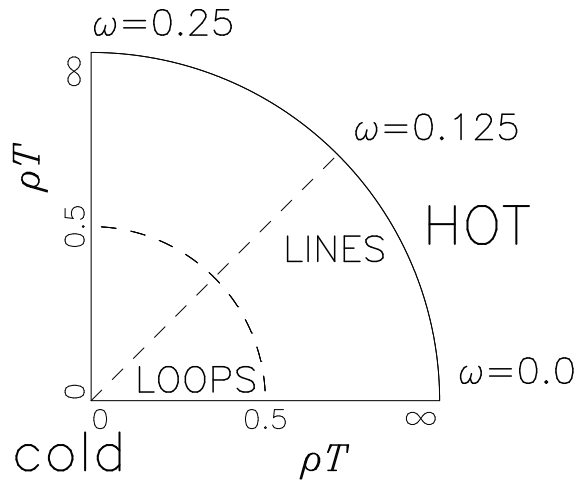}
\vspace{-0.4in}
\caption{Magnetic monopole phase plane for the caloron 
in polar co-ordinates: $(\rho T, 2 \pi \omega)$.}
\label{fig:phase}
\vskip -0.5cm
\end{figure}

The resultant picture that emerges is appealing (see
Fig. \ref{fig:phase}). For $\omega=0$, the
small monopole loop at the core of a cold instanton grows in size as
one increases the temperature and is transformed into a single static
't~Hooft-Polyakov monopole at infinite temperature, as noted earlier
by Rossi~\cite{ROSSI}. Note that the other quadrants of Fig. 1 can be found
by applying $Z_2$ center symmetry ($\omega \rightarrow \omega +
\half$) and monopole to anti-monopole charge conjugation ($\omega -
\quarter \rightarrow \quarter - \omega$).

The MA projection provides a fully gauge and Lorentz invariant
definition of monopole currents by introducing an auxiliary adjoint
Higgs field, $\vec \phi(x)$, fixed at the classical minimum,
\be
G = \half \int [ (D_\mu(A) \vec\phi)^2 + \lambda (\vec\phi \cdot\vec\phi
-1)^2 ] d^4x\;,
\ee
in a fixed background gauge field, $A_{\mu}$. This yields the
Abelian projected field strength,
\be 
f_{\mu \nu} = \n \cdot {\mathbf F}_{\mu\nu} - \n\cdot
D_\mu \n \times D_\nu \n \; ,
\ee 
with its  U(1) monopole current,
\be
k_\mu = \frac{1}{4 \pi} \d_\nu \tilde f_{\nu\mu}
\rightarrow  \frac{1}{8 \pi } \epsilon_{\mu\nu\rho\sigma} \d_\nu \n 
\cdot \d_\rho \n \times \d_\sigma \n \; ,
\ee
where $\n(x) \equiv \vec \phi(x)/|\phi(x)|$.  It is conventional to
identify the MA gauge by the rotation $\Omega(x)$ in the coset
$SU(2)/U(1)$ that aligns $\n$ along the 3 axis.  We  have
extended the conventional Abelian projection ($\lambda =
\infty$) to a continuous family including the analytically more
tractable BPS limit ($\lambda = 0$), where the difficult problem of
minimizing the MA functional $G$ reduces to an eigenvector problem for
the Higgs field, $D_\mu(A)^2 \vec \phi_E = E \vec \phi_E$.

\section{COLD MONOPOLE LOOP}

We begin at the origin of Fig. 1, where there is a
single isolated instanton at zero temperature.  A trivial, but
essential,  observation is that the singular gauge instanton
in the  't Hooft ansatz,
\be
A^a_\mu = \bar \eta^a_{\mu\nu}  \d_\nu \log(1 + \rho^2/x^2). 
\ee 
is {\bf also} the MA projection which minimizes the Higgs action
$G$. In the BPS limit, this is equivalent to having a 
zero eigenvalue solution, 
\[ \vec \phi_0(x) = \frac{x^2}{x^2 + \rho^2} [0,0,1] , \]
for the Higgs field aligned with the 3-axis. Consequently there is {\bf
no}  magnetic content to the MA projection.  This would be the
entire story except that there is another zero eigenvalue that
implies a flat direction for the formation of an infinitesimal
monopole loop.

The geometry of this loop is interesting. The gauge singularity at the
origin is caused by a rotation, $g(x) = x_\mu \tau_\mu/|x|$ and
$\tau_\mu = (1, i \vec \tau)$. Locally it is advantageous to
``unwind'' this singularity to a distance R further reducing the MA
functional $G$. This almost wins, creating a monopole loop of radius R
which only slightly increases G, $\delta G/G \simeq R^4
\log(R)$. Almost any local disturbance, due to a nearby instanton for
example, will stabilize the loop \cite{BOT}. The second zero mode
vector in the BPS limit is easily constructed using conformal
invariance.
\[
\vec \phi_1(x) = \frac{1}{x^2( x^2 + \rho^2)} [\sin \beta \cos \alpha, 
\sin \beta \sin \alpha, \cos \beta]  ,
\]
where $\alpha =\varphi - \psi$
, $\cos \beta = (v^2 - u^2)/(u^2 + v^2)\;$, with
$x_1 + ix_2 = u\E{i \varphi}$,  $x_3 + i x_0 = v\E{\psi}$. 

The superposition of the two zero modes produces a loop. Finally note
that the topological charge is related to the magnetic charge, through
a surface term on the boundary of the loop, $\Sigma$,
\[
Q  \rightarrow \int_\Sigma \frac{Tr[ \Omega_Rd\Omega_R\da
\wedge \Omega_Rd\Omega_R\da\wedge \Omega_Rd\Omega_R\da]}{24 \pi^2} \; ,
\]
where $\Omega_R$ is the singular gauge transformation providing a Hopf fibration for
the infinitesimal ``non-contractible'' loop.

\section{HOT BPS MONOPOLES}

At low temperature with $\omega = 0$ the MA projection
is similar. The periodic instanton in the 't Hooft ansatz,
\[
A^a_\mu \!\!=\! \bar \eta^a_{\mu\nu}  \d_\nu 
\log\!\left(\!1\! + \frac{\pi T \rho^2 \sinh( 2 \pi T r) }{2 r (
\sinh^2(\pi T r) + \sin^2(\pi T t))}\!\right)\! ,
\]
is again equivalent to the MA projection. ``Unwinding'' the periodic
copies of the singularities at $x = 0$ is now accomplished by $g =
X_\mu \tau_\mu/|X| \; , \; X_\mu = [\tan (\pi T t),\hat r \tanh( \pi T
r)]$, leaving a monopole loop. However surprisingly at infinite
temperature, or equivalently $\rho = \infty$ as noted by Rossi, the
instanton is gauge equivalent to the static 't Hooft-Polyakov monopole
solution. With $\n(x) = \hat r$, this is the correct MA projection (or
unitary gauge). For the case of the BPS limit, the solution is simply
$\vec \phi(x) \equiv \vec A_0(x)$, as one might expect. Consequently the MA
projection correctly identifies the standard static monopole.

\section{CALORON $\bf T$-$\bf \omega$ PLANE}

For $\omega\neq 0$, we encounter the full complexity of the new
caloron solution~\cite{KvB},
\[
A_\mu\!\! =\! \tau^3 \bar \eta^3_{\mu\nu}  \d_\nu \log\phi(x)
\!+\! (\tau^+ \bar \eta^-_{\mu\nu} \d_\nu \chi(x) \!+\! c.c.)
\psi(x),
\label{VB}
\]
in the singular gauge. In the limit of $T \rightarrow\infty$, we have
verified that MA projection now gives a pair of 't Hooft-Polyakov BPS
monopole/anti-monopole separated by distance $D = \pi \rho^2 T$ as
expected.  However, now the singular gauge caloron no longer satisfies
the MA projection and it is difficult to find the MA projection
analytically. Thus we have minimized G numerical in the interior of
the $T-\omega$ phase plane of Fig. 1 by placing the functional on a
grid.
\begin{figure}[t]
\vspace{-0.3in}
\epsfxsize=\hsize
\epsfysize=1.8in
\epsffile{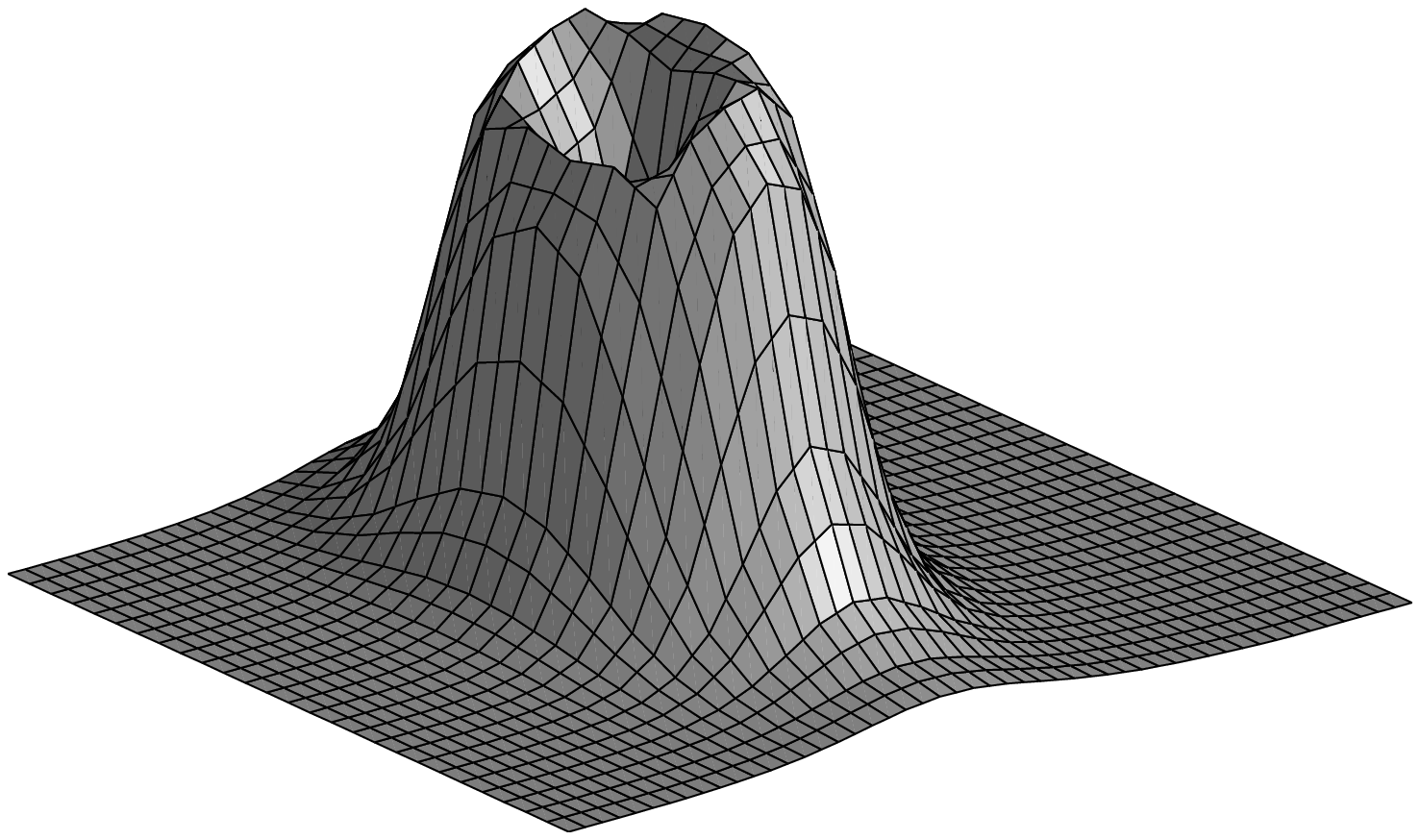}
\epsfxsize=\hsize
\epsfysize=1.8in
\epsffile{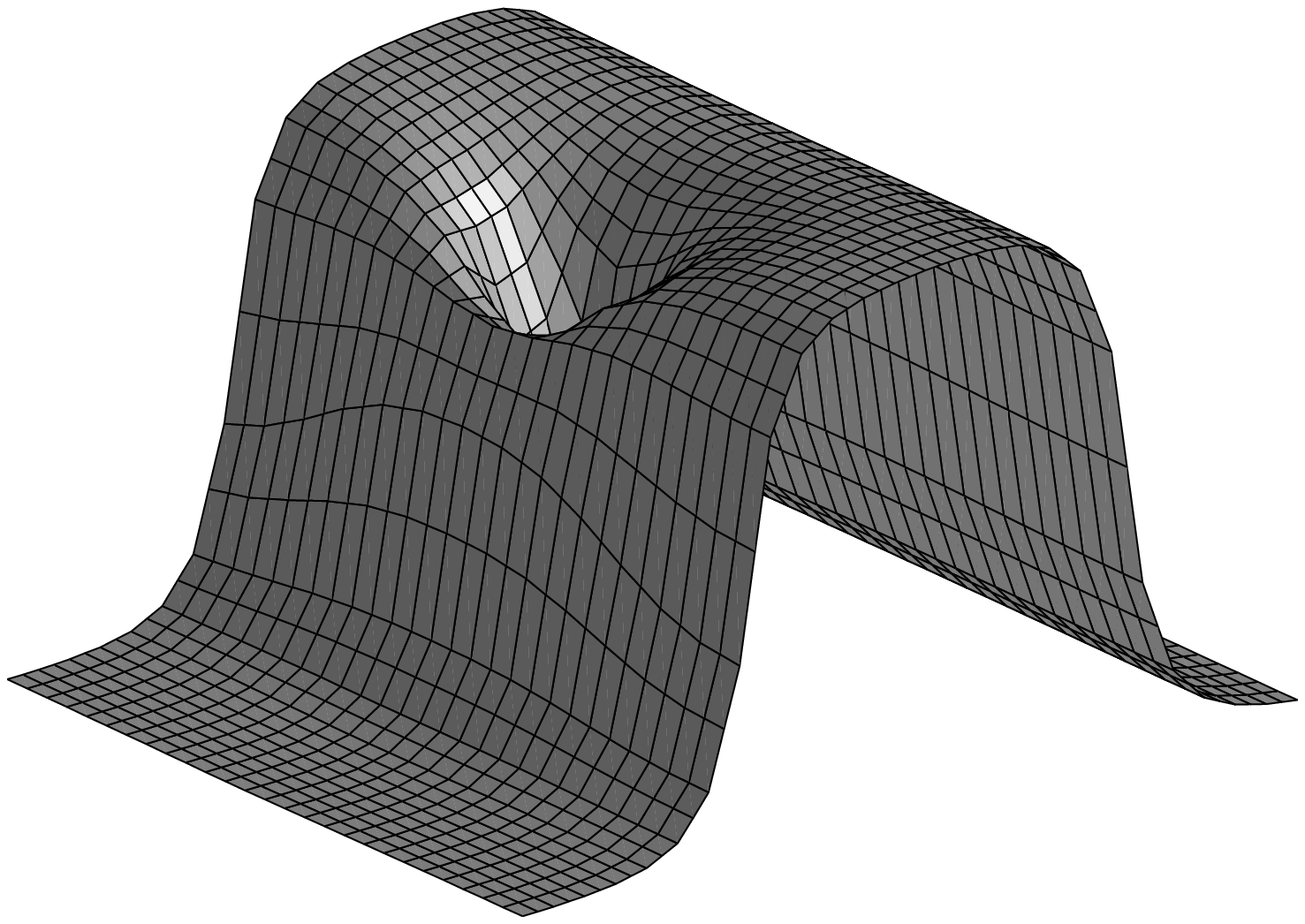}
\vspace{-0.4in}
\caption{Profiles of $\beta(0,z,t)$ at $\omega = .125$ 
for a loop (top, $\rho T = 0.56$) and 
for monopole/anti-monopole pair (bottom, $\rho T = 0.57$).}
\label{fig:line}
\vskip -1.0 cm
\end{figure}
On symmetry grounds, one can prove the existence of cylindrical
solutions with $\alpha =\varphi + \tilde \alpha(u,z,t) \;,\;
\beta(u,z,t)$. The Dirac sheet is located by a jump in $\beta(t,x)$ 
by $\pi$.  In Fig. 2, we give the profile for $\beta(u,z,t)$ in the z-t
plane slicing through the instanton centered at $u= 0$.
\begin{figure}[t]
\vspace{-0.2in}
\epsfxsize=\hsize
\epsffile{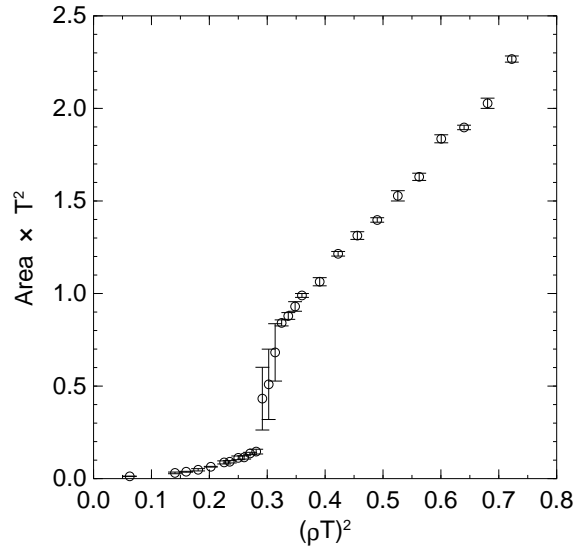}
\vspace{-0.4in}
\caption{The area $A$ of the minimal spanning Dirac  sheet versus 
temperature $T$ for  $\omega = 0.125$.}
\label{fig:dirac}
\vskip -0.5cm
\end{figure}

To explore further the transition from the monopole loop to a
pair of monopole lines, we plot in Fig. 3 the area of
the minimal spanning Dirac sheet. At $\rho T \simeq 0.56$, there is a
clear transition separating the two regimes.

Based on the absence of a loop for a single isolated
instanton~\cite{BOT}, we anticipate that the formula for the size
of the loop (or separation of the lines) must involve a new length
scale, $L$. This scale represents the distance to nearby perturbations
such as the anti-instanton presented in Ref.~\cite{BOT}. For the 
single caloron plotted here, the new scale is $L = \beta =1/T$ .  This suggests
a simple scaling form: $R \sim \rho (\rho T)^{\gamma}$ with
$\gamma>0$. Indeed in Fig. 3 for $(\rho T)^2 < 0.32$, we do see a
positive curvature for the area, $A/\beta^2 \sim (R T)^2 \sim (\rho
T)^{2+2\gamma}$, {\it i.e.}, $\gamma >0$, consistent with our
expectation.  On the other hand, at high temperatures,
the monopole/anti-monopole trajectories are known~\cite{KvB,LL} to be
separated asymptotically by $D=\pi \rho (\rho T)^\gamma$ with $\gamma
= 1$, which is also confirmed by a linear fit to $ A = D \beta
\simeq \pi \rho ^2$ for $(\rho T)^2 > 0.3$.

Finally it is interesting to note that the kinematical ``transition''
seen in Fig. 3 is near to the Yang-Mills deconfinement temperature
for a typical instanton size of $\smfrac{1}{3}$ fermi. However, a 
serious analysis of deconfinement dynamics and its possible relations
to the monopole content of the caloron is left to future
investigations.

\end{document}